# FIVE SUPERNOVA SURVEY GALAXIES IN THE SOUTHERN HEMISPHERE.
# II. THE SUPERNOVA RATES

A.A. HAKOBYAN[1,2], A.R. PETROSIAN[1], G.A. MAMON[3], B. MCLEAN[4], D. KUNTH[3],
M. TURATTO[5], E. CAPPELLARO[6], F. MANNUCCI[7], R.J. ALLEN[4], N. PANAGIA[4,8,9],
M. DELLA VALLE[10,11]

Based on the database compiled in the first article of this series, with 56 SN events discovered in 3838 galaxies of the southern hemisphere, we compute th rate of supernovae (SNe) of different types along the Hubble sequence normalized to the optical and near-infrared luminosities as well as to the stellar mass of the galaxies. We find that the rates of all SN types show a dependence on both morphology and colors of the galaxies, and therefore, on the star-formation activity. The rate of core-collapse (CC) SNe is confirmed to be closely related to the Star Formation Rate (SFR) and only indirectly to the total mass of the galaxies. The rate of SNe Ia can be explained by assuming that at least 15% of Ia events in spiral galaxies originates in relatively young stellar populations. We find that the rates show no modulation with nuclear activity or environment. The ratio of SN rates between types Ib/c and II shows no trend with spiral type.

**1. Introduction.** The rates of SNe of different types have two major roles in modern astrophysics. They are crucial parameters in the modeling of galaxy formation and chemical enrichment (e.g., [1,2]). Also, SNe rates are important tools to investigate the nature of the progenitor stars (e.g., [3-5]). The measured rates of SNe at low and at high redshifts are mainly based on optical observations. The rates in the local universe ($z < 0.1$) have been computed using the Five SN Survey (FSS) [6] data [3,7] as well as data of the still ongoing Lick Observatory Supernova Search (LOSS) project [8,9]. In order to obtain a direct measurement of the rates of SNe in a given galaxy sample, one has to divide the number of discovered SNe to the period of surveillance. The latter can be estimated by computing the *control time* through the detailed analysis of the log of a given SN search [10] after taking into account the various selection effects (e.g., [6,9]). In all cases, the SN rate must be normalized to some scaling quantity of the galaxy, which most often is its luminosity as an estimate of the stellar mass, particularly in the $B$ band (e.g., [6,11,12]), and recently also in the near- and far-infrared bands (e.g., [3,7,9,13]). Thus, the classical SN rate unit is defined as number of events per century per $10^{10} L_\odot$ in $B$ (SNuB) or in the near-infrared (SNuIR) bands.

It is well known that the $B$ band luminosity of galaxies, being the result of combined emission from old and young stellar populations and absorption by dust, changes dramatically along the Hubble sequence and cannot be used as a tracer of stellar mass (e.g., [14]). In contrast, the near-infrared (NIR) luminosity, in particular the *K* luminosity, is a better proxy for the stellar mass [14]. One can obtain an even better proxy for stellar mass by combining *K* luminosity with galaxy color [14]. The availability of multi-wavelength data extending to the NIR fluxes make possible more accurate determination of stellar masses of the galaxies and control of their dependence from the different stellar populations.

In the first article of this series [15], we presented and discussed the database [16] of 3838 galaxies that were monitored for SNe events. For this sample of galaxies, newly determined morphologies, angular sizes, inclinations, and fluxes in optical *U*, *B*, *R*, and NIR *I*, *J*, *H*, *K* bands as well as galaxy activity classes and information about their environment are available [16]. Using this database, we present here new estimates of the SN rates for different galaxy morphological types normalized to the different optical to NIR luminosities as well as to stellar mass of the galaxies. Our results, combined with multi-band photometry data, allow also to discuss the dependence of the rates of all SN types, i.e., Ia and core collapse (Ib/c and II, hereafter CC), on both the morphology and colors of the galaxies and, therefore, on the stellar populations and star formation activity.

Several authors have studied the possible connection between nuclear activity and/or starbursts in galaxies and the occurrence of SNe [17-21]. Roughly 18% of our 3838 galaxies are active or star-forming (A/SF) galaxies [15]. We can therefore use our sample to determine the rates of different types of SNe as a function of galaxy nuclear activity or bursting level of star-formation, and to study possible links between nuclear activity and SN explosions.

An additional issue is whether gravitational interaction is a possible star-formation triggering mechanism. Evidences in this respect come from statistical studies (e.g., [22]) as well as from several observations (e.g., [23]), in general indicating that most of star-formation takes place in the central regions of the galaxies (e.g., [24]). The relation between interaction and star-formation was also studied using SNe as tracers of star-formation. With the exception of strongly interacting systems, the SN host galaxy environment appears to have no direct influence on SN production [21], hence on star-formation. We can use measures of the local environment to determine the rates of different types of SNe as a function of the environment (e.g., [24-26]).

The outline of this paper is as follows. In Section 2 we summarize the SN input catalog and the galaxy database. We briefly illustrate our adopted method to compute the SN rates in Section 3. Our results are presented in Section 4 and discussed in Section 5. In this paper, we have assumed a value for the Hubble constant of $H_0 = 75$ km s$^{-1}$ Mpc$^{-1}$.

**2. *The SN and galaxy database.*** To properly determine the rates of various types of SNe normalized to the optical *U*, *B*, *R* and NIR *I*, *J*, *H*, *K* luminosities as well as to the stellar mass, we have created the database of FSS galaxies identified in the field of DEep Near Infrared Survey (DENIS) [16] of the southern sky. For each galaxy, we determined its morphology, major diameter, axial ratio, and position angle. All measurements were made using images extracted from the Space Telescope Science Institute (STScI) Digital Sky Survey (DSS) *I* (near-IR), *F* (red), and *J* (blue) band photographic plates obtained by the Palomar and UK Schmidt Telescopes (UKST). We also extracted *J*, *H*, $K_s$ magnitudes from the Two Micron All Sky Survey (2MASS) and *I*, *J*, *K* magnitudes from DENIS. In addition, to studying SN rate dependence on the activity level of the galaxies, a subsample of A/SF galaxies was created. Moreover, to study the dependence of the SN rate on the local environment of their hosts, the apparent number of neighboring galaxies within a 50 kpc radius was determined. The sample parameters are discussed in detail in [15]. In total our sample has 3838 FSS galaxies monitored for SNe, among which 56 SNe were detected within 53 galaxies.

The full FSS [7] monitored roughly 9700 galaxies, in both hemispheres, and discovered roughly 250 SNe. Because of missing data for some SNe and their host galaxies, only 136 SNe were used for the rate calculations [6,7]. The same problem occurs also in our study. From 3838 galaxies, identified in common FSS and DENIS fields, not all have the necessary data available to allow for SNe rate calculation. Table 1 summarizes the number of galaxies with all set of necessary parameters in different photometric bands, which were used for SNe rates calculations, as well as the number of discovered different type of SNe and their total number in each band. The non-integer numbers of SNe are due to the presence of a few SNe with incomplete or unknown classification which, as explained in [6], were divided into different classes. We assumed that all SNe with unknown classification in E or S0 galaxies belong to SN type Ia (e.g., [6,27]). This is justified by the absence of CC SNe among our E/S0s. On the other hand, SNe with unknown classifications in S0/a-Sd hosts have been redistributed among the three basic SN types according to the observed distribution of classified SNe in similar hosts (e.g., [6,7]).

*Table 1*

NUMBER OF SNe AND GALAXIES
PER PHOTOMETRIC BAND

| Band | $N_{gal}$ | $N_{Ia}$ | $N_{Ib/c}$ | $N_{II}$ | $N_{All}$ |
|---|---|---|---|---|---|
| *U* | 1079 | 12.4 | 6.4 | 18.2 | 37 |
| *B* | 3723 | 22.1 | 7.7 | 24.2 | 54 |
| *R* | 3696 | 22.1 | 7.7 | 24.2 | 54 |
| *I* | 3700 | 22.2 | 7.8 | 22.0 | 52 |
| *J* | 3338 | 22.1 | 7.7 | 24.2 | 54 |
| *H* | 3098 | 21.1 | 7.7 | 24.2 | 53 |
| *K* | 3262 | 21.1 | 7.7 | 24.2 | 53 |

Note: The fractional numbers arise from uncertain SN types.

**3.** *SN rate measurement method*. The SN rate computations were performed following the procedures and the assumptions used in [6] and [3,7], in fact using the same software as used in [6,7]. Normalizing SNe rates to luminosity in different bands is useful to sample different stellar populations and hence to obtain useful information for the progenitor scenarios. For example, some authors [28,29] normalized the rate of SN Ia to *H* and *K* luminosity: in these bands, the role of old stars in all galaxy types is dominant. They found a sharp increase of the production of type Ia SNe toward late morphological types. It is important to note that, in those computations, the normalization of the SN rate was carried out using average colors for a given class of the parent galaxies, because at that time direct and individual NIR measurements were not available for most galaxies. The multi-band photometric data of [15] can be used to determine the stellar population of the galaxies in a similar fashion as [28,29], but more accurately, as we use more photometric bands.

In the past, the luminosity in the *B* band has been mostly used as a proxy for the stellar mass of galaxies. The recent availability of NIR data has allowed a more accurate stellar mass determination. Using galaxy evolutionary synthesis models, Bell & de Jong [14] computed various linear relations $\log (M/L_\lambda) = a_\lambda + b_\lambda C_\lambda$ relating the stellar mass-to-light ratio (*M/L*) to the color *C*. The color dependence is smallest when an infrared band is used to define the color. These models can be used to compute the *M/L* ratio, hence stellar mass, along the whole Hubble sequence, from ellipticals to irregulars. From the coefficients provided in Table 1 of [14], the relation between the stellar mass, *K*-band luminosity ($L_K$) and the color (*B*−*K*) is

$$\log (M/L_K) = 0.212 (B-K) - 0.959 \quad . \tag{1}$$

Equation (1), which was also adopted in [3,8,9], was exploited to compute the stellar mass of each galaxy of our sample.

In this study we did not make a special effort to check our stellar mass determinations. We believe that the consistency check performed in [3] by comparing, for the total sample of FSS galaxies, the mass estimate using equation (1) with mass determinations using other different methods [30-32] is applicable also to our results.

**4. *Results.*** The SN rates computed in units of optical *U*, *B*, *R* and NIR *I*, *J*, *H*, *K* luminosities are shown in Table 2. The rates in this table are expressed in number of SNe per century per $10^{10}$ *U*, *B*, *R*, *I*, *J*, *H*, and *K* band solar luminosities (SNuU, SNuB, SNuK, etc.). Galaxy luminosities were derived from absolute magnitudes, assuming that the absolute magnitude of the Sun is 5.61, 5.48, 4.42, 4.08, 3.64, 3.32, and 3.28 [33] in our 7 bands, respectively.

Columns 1 and 2 respectively give the photometric band and morphology bin. Column 3 is the number of the galaxies in each photometric band and morphological bin. Columns 4, 5, 6, and 7 present the numbers of different types of SNe in the same intervals. Columns 8, 9, 10, and 11 give the SN rates normalized to the appropriate band luminosities. The rate errors combine the Poisson errors, often dominant, and uncertainties on the input parameters and bias corrections as explained in [6,7]. Upper limits were computed by assuming one SN in corresponding bins.

THE SN RATES NORMALIZED TO THE *U*, *B*, *R*, *I*, *J*, *H*, AND *K* LUMINOSITIES

| Band | Type | $N_{gal}$ | $N_{Ia}$ | $N_{Ib/c}$ | $N_{II}$ | $N_{All}$ | Rate [SNuBand] | | | |
|---|---|---|---|---|---|---|---|---|---|---|
| | | | | | | | Ia | Ib/c | II | All |
| (1) | (2) | (3) | (4) | (5) | (6) | (7) | (8) | (9) | (10) | (11) |
| *U* | E-S0 | 313 | 6.0 | 0.0 | 0.0 | 6.0 | 0.33±0.15 | <0.08 | <0.13 | 0.33±0.15 |
| | S0/a-Sb | 364 | 4.0 | 2.0 | 7.0 | 13.0 | 0.23±0.13 | 0.19±0.14 | 1.04±0.43 | 1.46±0.46 |
| | Sbc-Sd | 336 | 2.4 | 4.4 | 11.2 | 18.0 | 0.13±0.10 | 0.43±0.21 | 1.69±0.61 | 2.25±0.64 |
| | Sm-Irr | 66 | 0.0 | 0.0 | 0.0 | 0.0 | <1.41 | <2.06 | <3.18 | 0.00 |
| | All | 1079 | 12.4 | 6.4 | 18.2 | 37.0 | 0.23±0.10 | 0.19±0.09 | 0.84±0.32 | 1.16±0.31 |
| *B* | E-S0 | 930 | 8.0 | 0.0 | 0.0 | 8.0 | 0.21±0.08 | <0.04 | <0.07 | 0.21±0.08 |
| | S0/a-Sb | 1286 | 8.6 | 3.4 | 11.0 | 23.0 | 0.29±0.11 | 0.23±0.13 | 1.06±0.35 | 1.58±0.37 |
| | Sbc-Sd | 1300 | 5.5 | 4.3 | 13.2 | 23.0 | 0.24±0.12 | 0.40±0.20 | 1.78±0.59 | 2.42±0.60 |
| | Sm-Irr | 207 | 0.0 | 0.0 | 0.0 | 0.0 | <0.85 | <1.39 | <2.00 | 0.00 |
| | All | 3723 | 22.1 | 7.7 | 24.2 | 54.0 | 0.24±0.08 | 0.16±0.07 | 0.73±0.24 | 1.13±0.25 |
| *R* | E-S0 | 917 | 8.0 | 0.0 | 0.0 | 8.0 | 0.19±0.07 | <0.04 | <0.06 | 0.19±0.07 |
| | S0/a-Sb | 1280 | 8.6 | 3.4 | 11.0 | 23.0 | 0.29±0.11 | 0.24±0.14 | 1.05±0.35 | 1.57±0.37 |
| | Sbc-Sd | 1296 | 5.5 | 4.3 | 13.2 | 23.0 | 0.29±0.15 | 0.47±0.24 | 2.10±0.69 | 2.87±0.72 |
| | Sm-Irr | 203 | 0.0 | 0.0 | 0.0 | 0.0 | <1.17 | <1.95 | <2.75 | 0.00 |
| | All | 3696 | 22.1 | 7.7 | 24.2 | 54.0 | 0.24±0.08 | 0.16±0.07 | 0.70±0.23 | 1.10±0.24 |
| *I* | E-S0 | 924 | 8.0 | 0.0 | 0.0 | 8.0 | 0.15±0.06 | <0.03 | <0.05 | 0.15±0.06 |
| | S0/a-Sb | 1278 | 8.6 | 3.4 | 11.0 | 23.0 | 0.21±0.08 | 0.18±0.10 | 0.80±0.27 | 1.20±0.28 |
| | Sbc-Sd | 1293 | 5.6 | 4.4 | 11.0 | 21.0 | 0.28±0.14 | 0.55±0.27 | 1.85±0.67 | 2.68±0.70 |
| | Sm-Irr | 205 | 0.0 | 0.0 | 0.0 | 0.0 | <1.40 | <2.42 | <3.41 | 0.00 |
| | All | 3700 | 22.2 | 7.8 | 22.0 | 52.0 | 0.19±0.06 | 0.14±0.06 | 0.54±0.18 | 0.87±0.19 |
| *J* | E-S0 | 885 | 8.0 | 0.0 | 0.0 | 8.0 | 0.09±0.04 | <0.02 | <0.03 | 0.09±0.04 |
| | S0/a-Sb | 1225 | 8.6 | 3.4 | 11.0 | 23.0 | 0.15±0.06 | 0.13±0.07 | 0.57±0.19 | 0.85±0.20 |
| | Sbc-Sd | 1140 | 5.5 | 4.3 | 13.2 | 23.0 | 0.22±0.11 | 0.46±0.23 | 1.76±0.58 | 2.44±0.61 |
| | Sm-Irr | 88 | 0.0 | 0.0 | 0.0 | 0.0 | <1.87 | <3.81 | <4.90 | 0.00 |
| | All | 3338 | 22.1 | 7.7 | 24.2 | 54.0 | 0.13±0.04 | 0.09±0.04 | 0.40±0.13 | 0.62±0.14 |
| *H* | E-S0 | 844 | 8.0 | 0.0 | 0.0 | 8.0 | 0.07±0.03 | <0.02 | <0.02 | 0.07±0.03 |
| | S0/a-Sb | 1185 | 8.6 | 3.4 | 11.0 | 23.0 | 0.12±0.05 | 0.12±0.07 | 0.48±0.16 | 0.71±0.17 |
| | Sbc-Sd | 1017 | 4.5 | 4.3 | 13.2 | 22.0 | 0.14±0.08 | 0.37±0.19 | 1.35±0.45 | 1.85±0.47 |
| | Sm-Irr | 52 | 0.0 | 0.0 | 0.0 | 0.0 | <1.76 | <3.96 | <5.31 | 0.00 |
| | All | 3098 | 21.1 | 7.7 | 24.2 | 53.0 | 0.10±0.03 | 0.08±0.03 | 0.31±0.10 | 0.48±0.11 |
| *K* | E-S0 | 885 | 8.0 | 0.0 | 0.0 | 8.0 | 0.05±0.02 | <0.01 | <0.02 | 0.05±0.02 |
| | S0/a-Sb | 1226 | 8.6 | 3.4 | 11.0 | 23.0 | 0.08±0.03 | 0.07±0.04 | 0.32±0.11 | 0.47±0.11 |
| | Sbc-Sd | 1092 | 4.5 | 4.3 | 13.2 | 22.0 | 0.10±0.06 | 0.24±0.12 | 0.92±0.30 | 1.25±0.32 |
| | Sm-Irr | 59 | 0.0 | 0.0 | 0.0 | 0.0 | <1.24 | <3.06 | <3.73 | 0.00 |
| | All | 3262 | 21.1 | 7.7 | 24.2 | 53.0 | 0.07±0.02 | 0.05±0.02 | 0.22±0.07 | 0.34±0.07 |

In Table 3, the results of SN rates computations in unit of stellar mass are presented. They are expressed in number of SNe per century and per $10^{10} M_\odot$ stellar mass (SNuM).

For the rates (SNuB, SNuK and SNuM) as a function of galaxy Hubble type (particularly for the E-S0, S0/a-Sb, and Sbc-Sd classes) and *B−K* color, the results in [3,7,9] are generally in good agreement within the uncertainties. However, our rates are estimated for average galaxy sizes similar to [3,7] and we do not consider a Rate-Size Slope (RSS) as in [9]. We therefore mainly compare and discuss our rate results with those published in [3,7].

*Table 3*

THE SN RATES NORMALIZED TO THE STELLAR MASS

| Type | $N_{gal}$ | $N_{Ia}$ | $N_{Ib/c}$ | $N_{II}$ | $N_{All}$ | Rate [SNuM] | | | |
|---|---|---|---|---|---|---|---|---|---|
| | | | | | | Ia | Ib/c | II | All |
| (1) | (2) | (3) | (4) | (5) | (6) | (7) | (8) | (9) | (10) |
| E-S0 | 885 | 8.0 | 0.0 | 0.0 | 8.0 | 0.06±0.02 | < 0.02 | < 0.02 | 0.06±0.02 |
| S0/a-Sb | 1226 | 8.6 | 3.4 | 11.0 | 23.0 | 0.11±0.04 | 0.10±0.06 | 0.42±0.14 | 0.62±0.15 |
| Sbc-Sd | 1092 | 4.5 | 4.3 | 13.2 | 22.0 | 0.17±0.10 | 0.50±0.25 | 1.85±0.61 | 2.52±0.64 |
| Sm-Irr | 59 | 0.0 | 0.0 | 0.0 | 0.0 | < 2.57 | < 7.97 | < 8.88 | 0.00 |
| All | 3262 | 21.1 | 7.7 | 24.2 | 53.0 | 0.09±0.03 | 0.07±0.03 | 0.30±0.10 | 0.46±0.10 |

A significant (on average about 2.5 times) increase of the rates of SNe II and II+Ib/c (CC) from early (S0/a-Sb) to late-type (Sbc-Sd) spiral galaxies is obvious in all photometric bands (Table 2). This trend is stronger (about 4.7 times) when SNe rates are calibrated to the stellar masses (Table 3). Our data show that the *K*-normalized rate of CC SNe increases about 3.2 times from early to late-type spiral galaxies. This agrees well with the factor 3.6 increase of [3]. A similar trend of greater rates for later spiral types is also present for SNe Ib/c, but it is not statistically significant.

It is common understanding that all types of CC SNe (Ib, Ic, IIP, IIL, etc.) explode by the same mechanism. However, the observed optical spectra and light curves of CC SNe depend not only on the mechanism of explosion but also on the parameters of the progenitor star, particularly the mass of its H-rich envelope and its mass loss rate (e.g., [34]). The latter is mainly a function of the stellar mass and metallicity. It is then expected that the relative numbers of the various CC SNe, particularly Ib/c to II types should depend on the metallicity of the host galaxy (e.g., [35,36]). Thanks to the well-known dependencies of metallicity on stellar mass and SFR [37,38], the ratio of SNe Ib/c to II rates should also depend on galaxy luminosity, hence its morphology [36]. Following this reasoning we compared the ratios of the rates of SNe types Ib/c to II. We find that, in all colors, as well as in units of stellar mass, the rate of SNe Ib/c is on average a factor of ~4 lower than that of type II.

Because of their fundamental role in cosmology, the interest in the nature of SNe Ia and their progenitors (e.g., [39,40]) has dramatically increased. Tables 2 and 3 suggest that the rates of SNe Ia increase from E-S0 galaxies to Sbc-Sd spirals in all photometric bands (except *U*, Table 2), as well as for the case of calibration to the stellar mass (Table 3), as also previously found in *B* [7] and *K* [3], or for calibration to stellar mass [3]. However, none of these trends are statistically significant. The existence of a local maximum of the SNe Ia rate for early spirals in the *B*-band was also noted by [7]. Relatively smaller statistics of SN Ia in *U* may be responsible for the lower rate at this shorter wavelength.

It is well known that galaxy colors become bluer moving from early to late types and that this corresponds to a sequence in the specific SFR (SFR/stellar mass), which is about zero in ellipticals and maximum in late spirals. However, this is not a one-to-one relation, as the dust content and the star formation history introduce large spreads even for a given Hubble type. Nevertheless, galaxy broadband colors can be used as proxies for Specific Star Formation Rates (SSFRs). We have therefore computed the SN rates in units of stellar mass after binning the galaxies according to their $U-B$, $B-R$, $R-I$, $I-J$, $J-H$, $H-K$, and $B-K$ colors. To increase the number of objects in each color bin we add up Ib/c and II as CC SNe. The results are plotted in Fig.1.

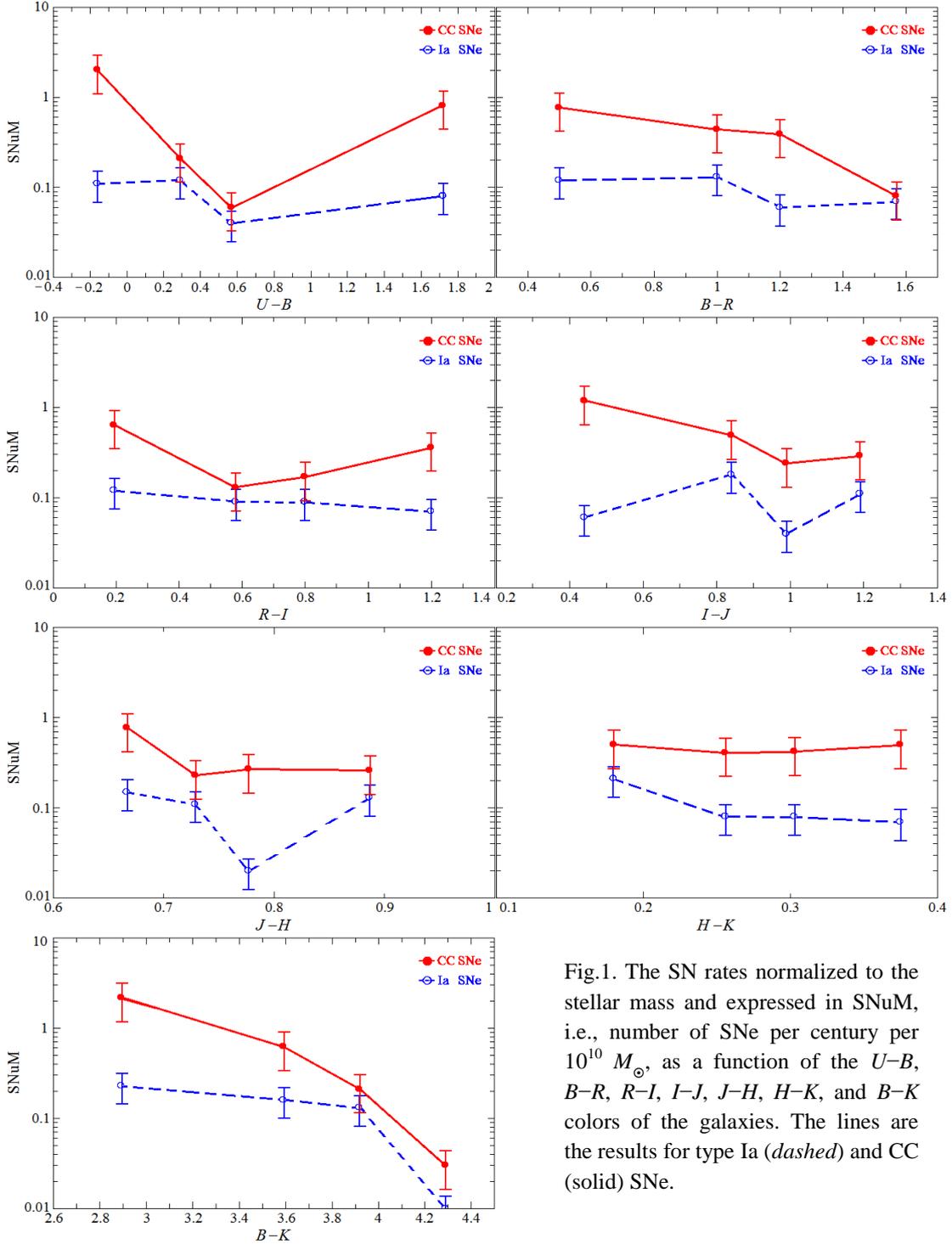

Fig.1. The SN rates normalized to the stellar mass and expressed in SNuM, i.e., number of SNe per century per $10^{10}$ $M_\odot$, as a function of the $U-B$, $B-R$, $R-I$, $I-J$, $J-H$, $H-K$, and $B-K$ colors of the galaxies. The lines are the results for type Ia (*dashed*) and CC (solid) SNe.

It is evident that the rates of CC SNe in the three colors ($U-B$, $B-R$, and $B-K$) that are sensitive SSFR indicators show a tendency to be higher in the bluer galaxies. A similar conclusion was obtained using $U-B$ and $U-V$ [7] and $B-K$ colors [3,9]. Instead in $R-I$, $I-J$, $J-H$, and $H-K$, which are less sensitive to SSFR, the rates of CC SNe within the uncertainties remain constant. Within the uncertainties, the SN Ia rate is independent on the host galaxy $R-I$, $I-J$, $J-H$, and $H-K$ colors (see also [7]). In $B-K$, the rate of SN Ia changes strongly with color, although not as quickly as for the CC SNe, again being higher in bluer galaxies. However, the increase in the rate of SNe Ia from red to blue galaxies is only significant for the reddest ones. In $U-B$ and $B-R$, this trend also exists but is not as strong. This result is in good agreement with the conclusion of [3].

SNe have occasionally been used as tracers of star-formation in galaxies with the aim to investigate to what extent nuclear starbursts or other nuclear activity may stimulate the star-formation activity in the host galaxy [21,41]. We address this question following the approach of [21,41] to specifically study the rates of SNe as a function of whether the host galaxy is classified as normal or whether it shows signs of activity such us a starburst or active nucleus. We computed the SN rates in units of stellar mass after binning the galaxies according to their activity level as explained in [15].

Table 4 shows our computed SN rates in units of stellar mass for normal as well as for A/SF galaxies. Comparing the rates of various types of SNe in normal and in A/SF galaxies, it is clear that in normal late-type spirals, in comparison with A/SF late spirals, the rate of II type SNe is significantly (about 4.5 times) higher. Lack of statistics prevents us from reaching a similar conclusion for types Ib/c and Ia SNe.

*Table 4*
THE SN RATES NORMALIZED TO THE STELLAR MASS IN NORMAL AND A/SF GALAXIES

| Subsample | Type | $N_{gal}$ | $N_{Ia}$ | $N_{Ib/c}$ | $N_{II}$ | $N_{All}$ | Rate [SNuM] | | | |
|---|---|---|---|---|---|---|---|---|---|---|
| | | | | | | | Ia | Ib/c | II | All |
| (1) | (2) | (3) | (4) | (5) | (6) | (7) | (8) | (9) | (10) | (11) |
| normal | E-S0 | 748 | 4.0 | 0.0 | 0.0 | 4.0 | 0.05±0.03 | <0.02 | <0.03 | 0.05±0.03 |
| | S0/a-Sb | 1002 | 6.8 | 1.2 | 5.0 | 13.0 | 0.15±0.07 | 0.08±0.07 | 0.45±0.22 | 0.68±0.21 |
| | Sbc-Sd | 901 | 3.6 | 1.2 | 11.2 | 16.0 | 0.20±0.13 | 0.22±0.21 | 3.05±1.09 | 3.48±1.04 |
| | Sm-Irr | 45 | 0.0 | 0.0 | 0.0 | 0.0 | <3.25 | <9.49 | <11.08 | 0.00 |
| | All | 2696 | 14.4 | 2.4 | 16.2 | 33.0 | 0.10±0.04 | 0.04±0.03 | 0.39±0.16 | 0.53±0.15 |
| A/SF | E-S0 | 137 | 4.0 | 0.0 | 0.0 | 4.0 | 0.09±0.05 | <0.04 | <0.07 | 0.09±0.05 |
| | S0/a-Sb | 224 | 2.0 | 2.0 | 6.0 | 10.0 | 0.06±0.05 | 0.11±0.08 | 0.46±0.21 | 0.63±0.23 |
| | Sbc-Sd | 191 | 1.0 | 3.0 | 2.0 | 6.0 | 0.13±0.13 | 0.87±0.52 | 0.68±0.58 | 1.67±0.82 |
| | Sm-Irr | 14 | 0.0 | 0.0 | 0.0 | 0.0 | <12.12 | <50.00 | <44.77 | 0.00 |
| | All | 566 | 7.0 | 5.0 | 8.0 | 20.0 | 0.08±0.05 | 0.11±0.06 | 0.24±0.14 | 0.43±0.15 |

We studied the dependence of SN rates on the density of the local environment. For each galaxy in our sample with redshift greater than 0.005, we have obtained counts within a circle of 50 kpc projected radius of neighboring galaxies with angular diameters within a factor 2 of the test galaxy [16]. We then subdivided our FSS galaxy sample into two subsamples: isolated galaxies, with no neighboring galaxy in a circle of 50 kpc projected radius ($n = 0$), and galaxies with at least one neighbor in the circle of the same projected radius ($n > 0$). In Table 5 we present the SNe rates in units of stellar mass for both subsamples. Comparing the rates of various types of SNe in galaxies with and without neighbor(s) we conclude that for all three types of SNe there are no significant differences between SNe rates.

*Table 5*

THE SN RATES NORMALIZED TO THE STELLAR MASS IN GALAXIES WITHOUT ANY NEIGHBORING OBJECT IN A CIRCLE OF 50 kpc RADIUS ($n = 0$), AND IN GALAXIES WITH ONE OR MORE NEIGHBORING OBJECT IN SAME RADIUS ($n > 0$)

| Subsample | Type | $N_{gal}$ | $N_{Ia}$ | $N_{Ib/c}$ | $N_{II}$ | $N_{All}$ | Rate [SNuM] | | | |
|---|---|---|---|---|---|---|---|---|---|---|
| | | | | | | | Ia | Ib/c | II | All |
| (1) | (2) | (3) | (4) | (5) | (6) | (7) | (8) | (9) | (10) | (11) |
| $n = 0$ | E-S0 | 496 | 5.0 | 0.0 | 0.0 | 5.0 | 0.07±0.03 | < 0.03 | < 0.04 | 0.07±0.03 |
| | S0/a-Sb | 808 | 5.0 | 2.0 | 7.0 | 14.0 | 0.09±0.05 | 0.08±0.06 | 0.36±0.15 | 0.52±0.16 |
| | Sbc-Sd | 837 | 3.4 | 4.4 | 10.2 | 18.0 | 0.16±0.11 | 0.59±0.29 | 1.70±0.64 | 2.45±0.69 |
| | Sm-Irr | 41 | 0.0 | 0.0 | 0.0 | 0.0 | < 5.40 | < 13.72 | < 15.81 | 0.00 |
| | All | 2182 | 13.4 | 6.4 | 17.2 | 37.0 | 0.09±0.04 | 0.09±0.04 | 0.33±0.13 | 0.51±0.13 |
| $n > 0$ | E-S0 | 389 | 3.0 | 0.0 | 0.0 | 3.0 | 0.05±0.03 | < 0.04 | < 0.05 | 0.05±0.03 |
| | S0/a-Sb | 418 | 3.8 | 1.2 | 4.0 | 9.0 | 0.16±0.09 | 0.14±0.13 | 0.61±0.34 | 0.92±0.35 |
| | Sbc-Sd | 255 | 1.0 | 0.0 | 3.0 | 4.0 | 0.23±0.23 | < 0.82 | 2.85±1.98 | 3.08±1.84 |
| | Sm-Irr | 18 | 0.0 | 0.0 | 0.0 | 0.0 | < 4.89 | < 19.05 | < 20.26 | 0.00 |
| | All | 1080 | 7.8 | 1.2 | 7.0 | 16.0 | 0.09±0.05 | 0.03±0.02 | 0.26±0.16 | 0.39±0.16 |

**5. *Discussion and conclusions.*** Using the database presented in the first article of this series [16], we compute the rates of SNe of different types grouping galaxies in four: E-S0, S0/a-Sb, Sbc-Sd, and Sm-Irr classes, and normalizing rates to the optical *U*, *B*, *R* and near-infrared *I*, *J*, *H*, *K* luminosities as well as to the stellar mass of the galaxies. We compute also the rate of SNe for samples of the galaxies with different level of nuclear activity and with different properties of their local environment. Close inspection of the results reveals the following:

1) The rate of type II SNe per unit of mass increases significantly (factor 4.4) from early to late-type spirals, in agreement with the understanding that rates of CC SNe are closely related to the SFR and only indirectly to the total stellar mass of the galaxies. A similar (factor 5) increase in the SNe Ib/c rates per unit mass from early to late-type spirals is present but not statistically significant. A similar tendency for the rate of CC SNe is visible in all photometric bands (see Table 2).

2) The rate of type Ia SNe per unit mass increases by a factor of about 2.8 from E-S0 to Sbc-Sd galaxies but is not statistically significant. A similar trend can be seen when the galaxies are binned according to their *U−B*, *B−R*, and *B−K* colors. In *B−K*, the ratio between SN Ia rates in galaxies bluer than *B−K* = 2.8 and redder than *B−K* = 4.2 is larger than 20, which is comparable with the values of ~30 times of [3] and ~15 times of [9]. The existence of such differences in type Ia SNe rates between late spirals and ellipticals implies that the frequency of progenitors exploding as a type Ia SN per unit time changes considerably with the ages of the parent population of the galaxies.

3) The key parameter relating SN Ia to the parent stellar population is the delay time distribution (DTD), i.e., the distribution of the time interval between the formation of the progenitor system and its explosion as a SN. Most binary star evolution models predict that these systems explode when the progenitor's age is between a few $10^7$ to $10^{10}$ years (e.g., [42]). Observational data support the predicted ages (e.g., [3,43,44]). Recent studies have suggested a possible bimodality of the DTD (e.g., [45]). Two formulations of the DTD bimodality were proposed. According to the first, so-called "weak" bimodality [46,47], 10% (5.5% in [47]) of the SNe Ia must explode on short timescales (~$10^8$ yrs) to follow the SFR, while the bulk of SNe Ia explodes on much longer timescales. According to the second, the so-called "strong" bimodality [45,48,49], a "prompt"-young component, comprising 20-60% of all the SNe Ia, explodes within $10^8$ yrs, while the "tardy"-old SNe explode on much longer timescale, up to Hubble time. In [3], a simple toy model was introduced in which the rate of type Ia SN is reproduced adding a constant contribution from "old" progenitors, independent of color and fixed at the value measured in the ellipticals, plus a contribution proportional to the rate of CC SNe. The best fitting agreement between observed SNuM = $f(B-K)$ and the toy model curves is obtained for the "young" progenitors fraction value of (35±8)% [3]. A similar analysis [9] produces a smaller (but statistically consistent) "young" progenitors fraction of (22±7)%. We have performed the same estimation using a slightly different approach. We assume that the SN Ia "old" progenitors belong to the bulge population of the galaxies and we fix the rate of SN Ia in bulges to that in galaxies with red $B$-$K$ colors (0.01 SNuM). We also assume that the "young" progenitors of SN Ia belong to the disk population of the galaxies. Taking the recent determination [50] of bulge to total mass ratios (B/T) for different types of galaxies, we estimated for the SN Ia "young" progenitors fraction a value of (15±7)% for early-type spirals, and (20±8)% for late-type spirals. This result is in agreement, within the errors, with the "prompt" fraction of SNe Ia previously reported. Figure 2 shows mass-weighted SN rate versus color and the toy model for SNe Ia.

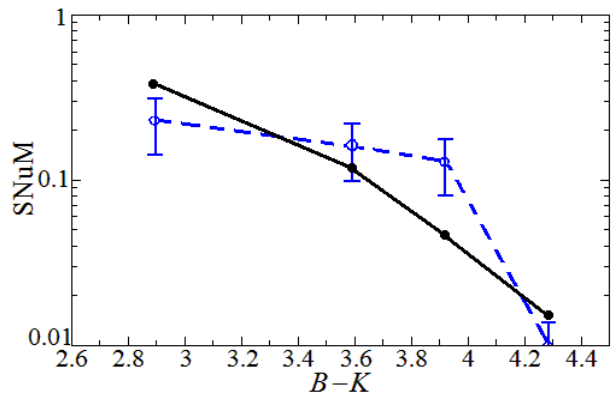

Fig.2. Mass-weighted SN rate versus color for SNe Ia. *Dashed*: data, *solid*: model.

4) We computed the SN rates in units of stellar mass after binning the galaxies according to their activity level (Table 4). Comparing the rates of various types of SNe in normal and in A/SF galaxies, we conclude that in general there is no significant difference between the rate of all SNe in normal (0.53±0.15) and in A/SF (0.43±0.15) galaxies. This conclusion holds also for CC SNe alone: there is no significant difference between CC SNe rates in normal (0.43±0.15) and in A/SF (0.35±0.14) galaxies. There is no significant difference between the rate of type Ia SNe in normal (0.10±0.04) and in A/SF (0.08±0.05) galaxies. Ratios of SN Ia to CC SN rates for normal and A/SF galaxies are the same and equal to ~0.23.

5) Using SNe as tracers of star-formation, we addressed also the problem of the relation between galaxies interaction and star-formation. For this, in Table 5, we presented SNe rates, in units of stellar mass, for galaxies without ($n = 0$), and with at least one neighboring object ($n > 0$). Comparing the rates of various types of SNe in galaxies without and with neighbor(s), we found that there is no significant difference between the rate of all SNe in galaxies with $n = 0$ (0.51±0.13) and with $n > 0$ (0.39±0.16) neighbors. This conclusion is confirmed separately for Ia and II types of SNe. Instead, Ib/c type SNe appear about 3 time more frequent in isolate ($n = 0$) galaxies in comparison with galaxies with neighbors ($n > 0$). This result reflects on the CC SNe rates. CC SNe in isolated galaxies are about 1.5 times more frequent than in galaxies with neighbors, and SNe Ib/c to SNe II rates ratio is more than 2 times higher in $n = 0$ (~0.27) than in $n > 0$ galaxies (~0.12). Again, we stress that this result is uncertain due to small statistics, but it is intriguing that galaxies with $n > 0$ usually are located in clusters where old stellar population members are expected to be dominant. The finding that Ia to CC SNe rates ratio in $n > 0$ galaxies (~0.31) is about 1.5 times higher than in $n = 0$ galaxies (~0.21) seems consistent with the predictions.

In a forthcoming paper, our results will be improved with the addition of the best current SN database, the LOSS one, which is the world's most successful search engine for nearby SNe. Its most recent database is based on a homogeneous set of several hundred SNe detected in ten years of CCD searches. Also, we are in the process of obtaining large and well-defined sample of targeted galaxies using the recent databases, such as 2MASS, Sloan Digital Sky Survey (SDSS), and Galaxy Evolution Explorer (GALEX). This would lead to obtaining the full Spectral Energy Distribution (SED) of the galaxies from near-UV to NIR, computing the Star Formation History (SFH) for each galaxy, and computing the DTD of SNe.

With the current data of the estimated SN rate, it is difficult to discriminate between different DTDs and then between different SN Ia progenitor models. To continue the study of SNe Ia rates in different stellar populations, observations at submillimeter and radio would be very useful. Measurements of the SN Ia rate in star forming and in passively evolving galaxies over a wide range of redshifts can provide more significant evidence about SN Ia progenitors and more detailed analysis of the role of SNe Ia in the metal enrichment.

*Acknowledgments.* A.A.H. and A.R.P. acknowledge the hospitality of the Institut d' Astrophysique de Paris (Paris, France) during their stay as visiting scientists supported by the Collaborative Bilateral Research Project of the State Committee of Science (SCS) of the Republic of Armenia and the French Centre National de la Recherché Scientifique (CNRS). A.R.P. acknowledges also the hospitality of the Space Telescope Science Institute (Baltimore, USA) during his stay as visiting scientist supported by the Director's Discretionary Research Fund. This work was made possible in part by a research grant from the Armenian National Science and Education Fund (ANSEF) based in New York, USA.

[1] Byurakan Astrophysical Observatory, 0213 Byurakan, Aragatsotn Province, Armenia, e-mail: hakobyan@bao.sci.am
[2] Yerevan State University, 1 Alex Manoogian, 0025 Yerevan, Armenia
[3] Institut d'Astrophysique de Paris (UMR 7095: CNRS & UPMC), 98 bis Bd Arago, 75014 Paris, France
[4] Space Telescope Science Institute, 3700 San Martin Drive, Baltimore, MD 21218, USA
[5] INAF - Osservatorio Astronomico di Trieste, Via Tiepolo 11, 34143 Trieste, Italy
[6] INAF - Osservatorio Astronomico di Padova, Vicolo dell'Osservatorio 5, 35122 Padova, Italy
[7] INAF - Osservatorio Astrofisico di Arcetri, Largo E. Fermi 5, 50125 Firenze, Italy
[8] INAF - Osservatorio Astrofisico di Catania, Via Santa Sofia 78, 95123 Catania, Italy
[9] Supernova Ltd., OYV #131, Northsound Road, Virgin Gorda, British Virgin Islands
[10] INAF - Osservatorio Astronomico di Capodimonte, Salita Moiariello 16, 80131 Napoli, Italy
[11] International Center for Relativistic Astrophysics, Piazzale della Repubblica 2, 65122 Pescara, Italy